\documentclass[aps,prl,twocolumn,superscriptaddress,showpacs,preprintnumbers,byrevtex,floatfix]{revtex4}
\usepackage{graphicx} 
\usepackage{amsmath}
\usepackage{array}
\usepackage{subfigure} 
\usepackage{epsfig}
\usepackage{url}   
\usepackage{dcolumn}  
\usepackage{color}

\graphicspath{{ps}}

\newcommand{\ks}{K_S^0} 
 
\newcommand{\bn}{B^0} 
 
\newcommand{\bbbar}{B\overline{B}} 
\newcommand{\ksks}{K_S^0 K_S^0}
\newcommand{\aksks}{{\cal A}_{\ksks}}
\newcommand{\sksks}{{\cal S}_{\ksks}}
\newcommand{\acal}{{\cal A}}
\newcommand{\scal}{{\cal S}}

\newcommand{\totbb}{657}
\newcommand{\mbc}{M_{\rm bc}}
\newcommand{\de}{\Delta E}
\newcommand{\qq}{q\overline{q}}
\newcommand{\qbarq}{\overline{q}q}
\newcommand{\Lsb}{\cal R}
\newcommand{\Ls}{\cal L_{\rm sig}}
\newcommand{\Lb}{\cal L_{\rm bkg}}
\newcommand{\dt}{\Delta t}
\newcommand{\bksks}{B^0 \to \ksks} 
\newcommand{\bkspc}{B^{+}\to \ks\pi^{+}} 
\newcommand{\avalue}{-0.38}
\newcommand{\astaterr}{\pm0.38}
\newcommand{\asysterr}{\pm0.05}
\newcommand{\svalue}{-0.38}

\newcommand{\sstaterrminus}{-0.77}
\newcommand{\sstaterrplus}{+0.69}
\newcommand{\ssysterr}{\pm0.09}

\newcommand{\jpsiks}{J / \psi \ks}

\begin{document}


\preprint{\vbox{ \hbox{   }
                 \hbox{BELLE Preprint 2007-51}
                 \hbox{KEK Preprint 2007-69}
}}

\title{ \quad\\[1.0cm] Measurement of Time-Dependent $CP$-Violating Parameters in $\bksks$ decays}


\affiliation{Budker Institute of Nuclear Physics, Novosibirsk}
\affiliation{Chiba University, Chiba}
\affiliation{University of Cincinnati, Cincinnati, Ohio 45221}
\affiliation{The Graduate University for Advanced Studies, Hayama}
\affiliation{Hanyang University, Seoul}
\affiliation{University of Hawaii, Honolulu, Hawaii 96822}
\affiliation{High Energy Accelerator Research Organization (KEK), Tsukuba}
\affiliation{University of Illinois at Urbana-Champaign, Urbana, Illinois 61801}
\affiliation{Institute of High Energy Physics, Chinese Academy of Sciences, Beijing}
\affiliation{Institute of High Energy Physics, Vienna}
\affiliation{Institute of High Energy Physics, Protvino}
\affiliation{Institute for Theoretical and Experimental Physics, Moscow}
\affiliation{J. Stefan Institute, Ljubljana}
\affiliation{Kanagawa University, Yokohama}
\affiliation{Korea University, Seoul}
\affiliation{Kyungpook National University, Taegu}
\affiliation{\'Ecole Polytechnique F\'ed\'erale de Lausanne (EPFL), Lausanne}
\affiliation{Faculty of Mathematics and Physics, University of Ljubljana, Ljubljana}
\affiliation{University of Maribor, Maribor}
\affiliation{University of Melbourne, School of Physics, Victoria 3010}
\affiliation{Nagoya University, Nagoya}
\affiliation{Nara Women's University, Nara}
\affiliation{National Central University, Chung-li}
\affiliation{National United University, Miao Li}
\affiliation{Department of Physics, National Taiwan University, Taipei}
\affiliation{H. Niewodniczanski Institute of Nuclear Physics, Krakow}
\affiliation{Nippon Dental University, Niigata}
\affiliation{Niigata University, Niigata}
\affiliation{University of Nova Gorica, Nova Gorica}
\affiliation{Osaka City University, Osaka}
\affiliation{Osaka University, Osaka}
\affiliation{Panjab University, Chandigarh}
\affiliation{RIKEN BNL Research Center, Upton, New York 11973}
\affiliation{Saga University, Saga}
\affiliation{University of Science and Technology of China, Hefei}
\affiliation{Seoul National University, Seoul}
\affiliation{Sungkyunkwan University, Suwon}
\affiliation{University of Sydney, Sydney, New South Wales}
\affiliation{Toho University, Funabashi}
\affiliation{Tohoku Gakuin University, Tagajo}
\affiliation{Department of Physics, University of Tokyo, Tokyo}
\affiliation{Tokyo Institute of Technology, Tokyo}
\affiliation{Tokyo Metropolitan University, Tokyo}
\affiliation{Tokyo University of Agriculture and Technology, Tokyo}
\affiliation{Virginia Polytechnic Institute and State University, Blacksburg, Virginia 24061}
\affiliation{Yonsei University, Seoul}
   \author{Y.~Nakahama}\affiliation{Department of Physics, University of Tokyo, Tokyo} 
   \author{K.~Sumisawa}\affiliation{High Energy Accelerator Research Organization (KEK), Tsukuba} 
   \author{I.~Adachi}\affiliation{High Energy Accelerator Research Organization (KEK), Tsukuba} 
   \author{H.~Aihara}\affiliation{Department of Physics, University of Tokyo, Tokyo} 
   \author{T.~Aushev}\affiliation{\'Ecole Polytechnique F\'ed\'erale de Lausanne (EPFL), Lausanne}\affiliation{Institute for Theoretical and Experimental Physics, Moscow} 
   \author{A.~M.~Bakich}\affiliation{University of Sydney, Sydney, New South Wales} 
   \author{V.~Balagura}\affiliation{Institute for Theoretical and Experimental Physics, Moscow} 
   \author{E.~Barberio}\affiliation{University of Melbourne, School of Physics, Victoria 3010} 
   \author{I.~Bedny}\affiliation{Budker Institute of Nuclear Physics, Novosibirsk} 
   \author{K.~Belous}\affiliation{Institute of High Energy Physics, Protvino} 
% ...
\collaboration{The Belle Collaboration}
\noaffiliation

\begin{abstract}
We report a measurement of the $CP$-violating parameters in $\bksks$ decays based on a data sample of $\totbb$$\times 10^6$ $\bbbar$ pairs collected at the $\Upsilon(4S)$ resonance with the Belle detector at the KEKB asymmetric-energy $e^+e^-$ collider.
In this study, one neutral $B$ meson is fully reconstructed in the $\bksks$ decay mode, and the flavor of the accompanying $B$ meson is identified by its decay products. The $CP$-violating parameters are measured from the asymmetry in the distributions of the proper-time interval between the two $B$ decays: $\sksks = \svalue ^{\sstaterrplus} _{\sstaterrminus}({\rm stat})\ssysterr ({\rm syst})$ and $\aksks = \avalue \astaterr ({\rm stat})\asysterr ({\rm syst})$. 
\end{abstract}


\maketitle

\tighten

{\renewcommand{\thefootnote}{\fnsymbol{footnote}}}

\setcounter{footnote}{0}
$\bn$ meson decays proceeding via flavor-changing $b \to d\qbarq$ or $s\qbarq$ transitions are sensitive
to new physics (NP) contributions affecting the internal quark loop diagrams.
Such NP contributions can add new weak phases and subsequently cause deviations
from the standard model (SM) expectations for $CP$-violating parameters~\cite{b2sTheory}.

The decay $\bksks$ is dominated by $b \to d\overline{s}s$ transitions.
The SM predicts that the $CP$-violating parameters $\sksks$ and $\aksks$ are zero
in the limit that the top quark dominates
and the contribution from internal up and charm quark exchanges in the loop diagram is small~\cite{ksksTheory1}.
Measurements of $\sksks$ and $\aksks$ are sensitive probes of NP~\cite{ksksTheory2}.
The parameter $\sksks$ arises from interference between a mixing-induced amplitude and a non-mixed decay amplitude
while the parameter $\aksks$ arises from $CP$ violation in the decay amplitude itself.
Experimentally, there are no prompt charged tracks from the $B$ vertex for a $\bksks$ decay,
and hence the $B$ decay vertex must be reconstructed by using $\ks$'s that decay
inside the vertex detector and a constraint on the beam interaction point (IP).
Both $CP$-violating parameters have previously been measured by the BaBar collaboration;
they obtained a large $\sksks$ value, albeit with a large statistical error~\cite{ksksBabar}.
In this Letter, we report a measurement of $\sksks$ and $\aksks$ in $\bksks$ decays
using almost twice the statistics of the previous measurement.
Our analysis is based on $\totbb$$\times 10^6$ $\bbbar$ pairs collected with the Belle detector~\cite{Belle} running at the KEKB asymmetric-energy $e^+e^-$ (3.5 on 8~GeV) collider~\cite{KEKB}.

In the decay chain $\Upsilon(4S)\to B\overline{B} \to (\ksks)f_{\rm tag}$, where one of the $B$ mesons decays
at time $t_{\ksks}$ to $\ksks$ and the other decays at time $t_{\rm tag}$ to a flavor specific state $f_{\rm tag}$
that distinguishes between $B^0$ and $\overline{B}{}^0$, the decay rate has a time dependence~\cite{Sanda} given by
\begin{eqnarray}
{\cal P}_{\ksks}(\Delta t)
= \frac{e^{-|\Delta t|/\tau_{B^0}}}{4\tau_{B^0}}
[1 + q \cdot \{ \sksks \sin(\Delta m_d \Delta t) \nonumber \\
            + \aksks \cos(\Delta m_d \Delta t ) \} ],
\label{eq_signal_pdf}
\end{eqnarray}
where $\Delta t = t_{\ksks} - t_{\rm tag}$,
$\tau_{B^0}$ is the $B^0$ lifetime,
$\Delta m_d$ is the mass difference between the two $B$ mass eigenstates,
and $q=+1~(-1)$ for $f_{\rm tag} = B^0~(\overline{B}{}^0)$.

At the KEKB, the $\Upsilon(4S)$ resonance is produced with a Lorentz boost of $\beta\gamma$ = 0.425
nearly along the $+z$ axis, which is defined as the direction antiparallel to the $e^+$ beamline.
Since the $B^0$ and $\overline{B}{}^0$ mesons are approximately at rest
in the $\Upsilon(4S)$ center-of-mass system (cms),
$\dt$ can be determined from the displacement in $z$ between the $\ksks$ and $f_{\rm tag}$ decay vertices:
$\dt \simeq$ ($z_{\ksks}-z_{\rm tag}$)/($\beta \gamma c$) $\equiv$ $\Delta z$/($\beta \gamma c$).

The Belle detector
is a large-solid-angle magnetic
spectrometer that
consists of a silicon vertex detector (SVD),
a 50-layer central drift chamber, an array of
aerogel threshold Cherenkov counters,
a barrel-like arrangement of time-of-flight
scintillation counters, an electromagnetic calorimeter,
which are located inside a superconducting solenoid coil that provides a 1.5~T
magnetic field. An iron flux return located outside of
the coil is instrumented to detect $K_L^0$ mesons and to identify muons.
Two inner detector configurations were used. A 2.0 cm-radius beam pipe
and a 3-layer SVD (SVD1) was used for the first data sample
of 152$\times 10^6$ $\bbbar$ pairs, while a 1.5 cm-radius beam pipe, a 4-layer
SVD (SVD2)~\cite{svd2}, and a small-cell inner drift chamber were used to record
the remaining 505$\times 10^6$ $\bbbar$ pairs.

We reconstruct a $\ks \to \pi^+ \pi^-$ candidate from a pair of oppositely charged tracks
having $|\Delta M_{\ks}| <$ 0.015 GeV/$c^2$ corresponding to three standard deviations ($\sigma$),
where $\Delta M_{\ks}$ is the difference between their invariant mass and the nominal $\ks$ mass~\cite{PDG}.
Both charged tracks are required to be displaced from the IP in the transverse ($r$-$\phi$) plane
by more than 100 $\mu$m.
The angle in the transverse plane between the $\ks$ momentum vector and the direction defined by the $\ks$ vertex
and the IP should be less than 50 mrad.
In order to suppress incorrect combinations of the two charged tracks,
the mismatch in the $z$ direction at the $\ks$ vertex point for the two charged tracks
is required to be less than 15 cm.

To identify $\bksks$ decay candidates, we use two kinematic variables:
the energy difference $\de \equiv E_B^{\rm cms} - E_{\rm beam}^{\rm cms}$
and the beam-energy constrained mass $\mbc \equiv \sqrt{(E_{\rm beam}^{\rm cms})^2 - (p_B^{\rm cms})^2}$,
where $E_{\rm beam}^{\rm cms}$ is the beam energy in the cms and $E_{B}^{\rm cms}$
and $p_{B}^{\rm cms}$ are the cms energy and momentum, respectively, of reconstructed $B$ candidates.
We select candidates satisfying $|\de|<0.20~{\rm GeV}$ and $5.20~{\rm GeV}/c^2<\mbc<5.30~{\rm GeV}/c^2$.
For the $\dt$ fit described below, we use candidates in a signal region defined as $|\de|<0.10~{\rm GeV}$
and $5.27~{\rm GeV}/c^2<\mbc<5.30~{\rm GeV}/c^2$.
We find that 0.2$\%$ of the selected events have multiple $\bksks$ candidates.
In such events, we choose the $\bksks$ candidate having the smallest $\Sigma$($\Delta M_{\ks})^2$ value.

To suppress continuum $e^+e^- \to q\bar{q}$ ($q$ = $u$,$d$,$s$,$c$) events,
we form a likelihood $\Ls$ ($\Lb$) for signal (continuum) events
by combining a Fisher discriminant based on modified Fox-Wolfram moments~\cite{SFW}
with the probability density function (PDF) for the cosine of the cms $\bn$ flight direction
with respect to the $+z$ axis. The former makes use of the difference in event shapes:
signal events have a spherical topology while background events tend to be jet-like.
We impose a requirement on the likelihood ratio $\Lsb = \Ls/(\Ls+\Lb)$
that retains 89$\%$ of the signal and rejects 71$\%$ of the continuum.

The $b$-flavor of the accompanying $B$ meson is identified from inclusive properties of particles
that are not associated with the reconstructed $\bksks$ candidate.
The tagging information is represented by two parameters: $q$, as defined in Eq.~(\ref{eq_signal_pdf}), and $r$,
which is an event-by-event Monte-Carlo-determined flavor-tagging dilution factor
that ranges from $r$~=~0 for no flavor discrimination to $r$~=~1
for unambiguous flavor assignment~\cite{TaggingNIM}.
Candidate events are selected to have $r>$~0.1, and are further divided into six $r$ intervals.
The wrong tag fraction $w$ for each $r$ interval and the differences $\Delta w$
between $\bn$ and $\overline{B}{}^0$ decays are determined
using semileptonic and hadronic $b \to c$ decay data~\cite{TaggingNIM}.

The dominant background is continuum.
We find the $\bbbar$ decay background contribution to be negligibly small
using a large sample of GEANT-based Monte Carlo (MC) simulated events~\cite{MC}.
Thus, we take into account only signal and continuum events in the nominal fit.
The uncertainty due to a possible contribution from $\bbbar$ decay background is included in the systematic errors.

\begin{figure}[htb]
\begin{center}
\includegraphics[height=150pt,width=!]{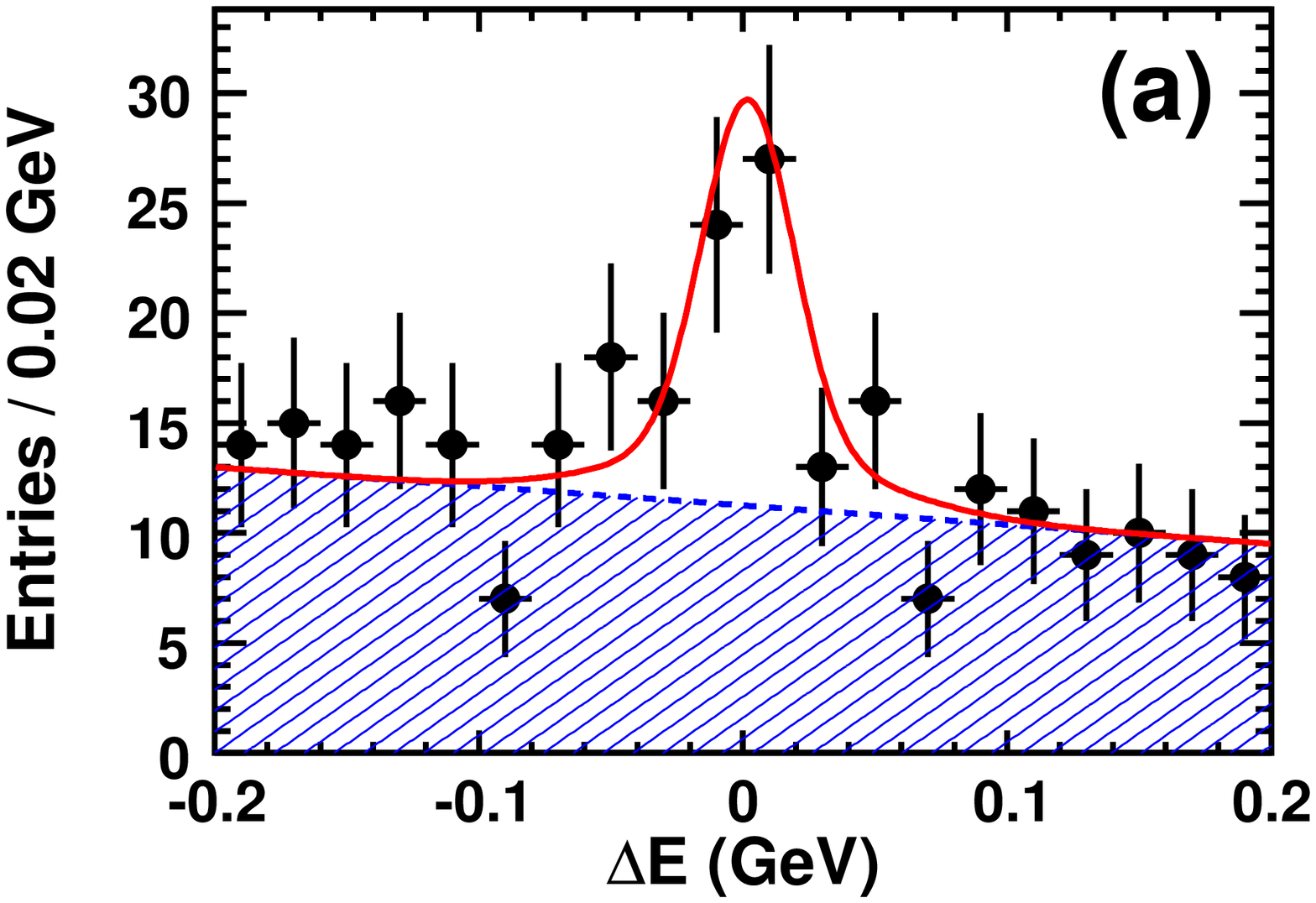} 
\includegraphics[height=150pt,width=!]{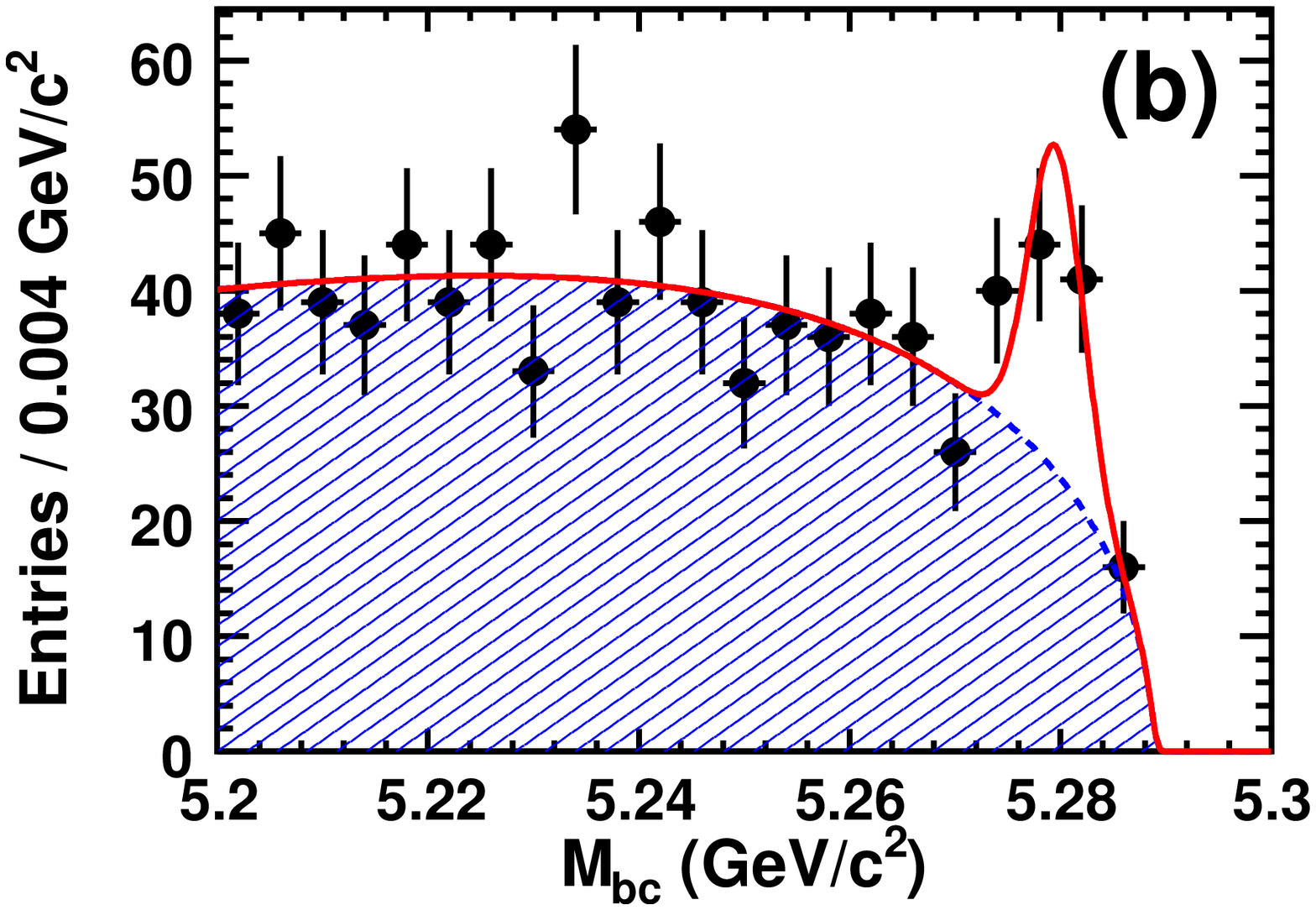}
\includegraphics[height=150pt,width=!]{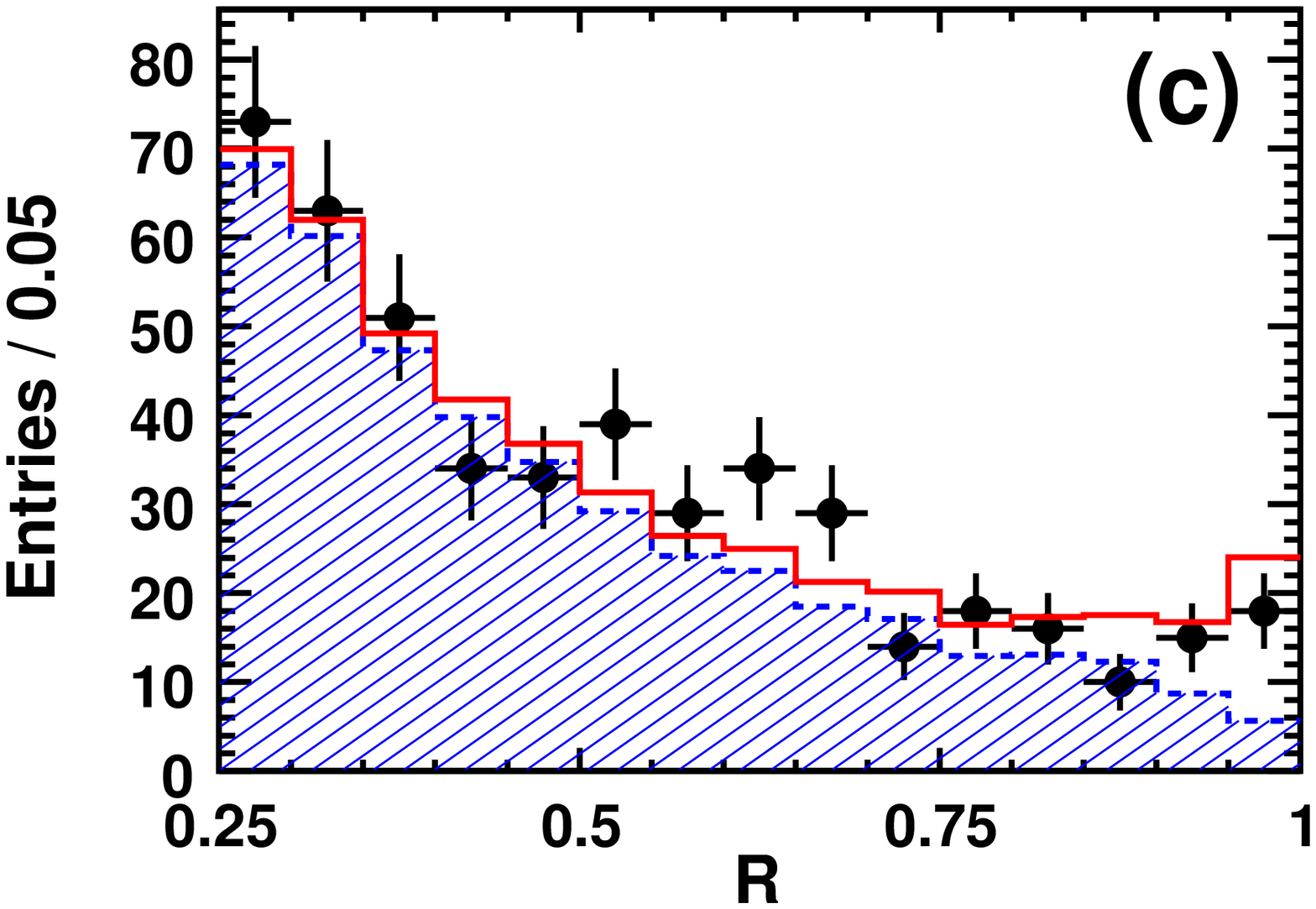}  
\caption{
(a) $\de$, (b) $\mbc$ and (c) $\Lsb$ projections for the $\bksks$ candidate events
(a) with $\Lsb >$~0.6 and $5.27~{\rm GeV}/c^2<\mbc$,
(b) with $\Lsb >$~0.6 and $|\de|<0.1~{\rm GeV}$ and
(c) in the signal region.
The solid histogram and curves show the fit projections and the hatched areas show the background component.
The points with error bars are the data.}
\label{fig_yield}
\end{center}
\end{figure}

The signal yield is extracted using a three-dimensional extended unbinned maximum likelihood (UML)
fit to $\de$-$\mbc$-$\Lsb$ distributions for the selected candidate events.
For the signal component,
we model the $\de$ ($\mbc$) shape using a sum of two Gaussians (a single Gaussian).
A binned histogram is employed for the $\Lsb$ distribution.
The parameters of the Gaussians and $\Lsb$ distribution are obtained using MC simulation.
For the background component,
the $\de$~($\mbc$) shape is modeled as a first-order polynomial (an ARGUS~\cite{argus}) function.
The parameters of these functions are floated in the fit.
The $\Lsb$ background distribution is obtained from a data sample in the sideband region~($\mbc <$~5.26~GeV/$c^2$).
Possible correlations among $\de$, $\mbc$ and $\Lsb$ are found to be negligible for signal events from the signal MC,
and to be very small for continuum events from the data sideband.
We include the effect of the small correlations in the latter in the systematic errors.
The fit yields 58$\pm$11 signal events among 476 $\bksks$ candidate events in the signal region,
where the error is statistical only. Figure~\ref{fig_yield} shows
the projections of the $\de$, $\mbc$ and $\Lsb$ distributions for the candidate events.

We apply the $B$ decay vertex reconstruction algorithm of Ref.~\cite{ksvertex}.
The vertex position for a $\bksks$ decay is obtained using the $\ks \to \pi^+\pi^-$ momentum vector
and a constraint on the IP;
the IP profile ($\sigma _x \simeq 100 \mu$m, $\sigma _y \simeq 5\mu$m)
is smeared by the finite $B^0$ flight length in the plane perpendicular to the $z$ axis.
To reconstruct the $\bksks$ decay position, both charged pions from at least one of the $\ks$'s
are required to have a sufficient number of hits in the SVD:
at least one layer with hits on both the $z$ and $r$-$\phi$ sides
and at least one additional layer with a hit on the $z$ side among the other layers for SVD1,
and at least two layers with hits on both sides for SVD2.
Both $\ks$'s are used for the vertex reconstruction if the four pions have a sufficient number of hits in the SVD.
The typical vertex reconstruction efficiency with SVD1 (SVD2) is determined to be 44$\%$ (61$\%$) from the signal MC.
The vertex position resolutions in the $z$ direction with SVD1 (SVD2) are 73$\mu$m (105$\mu$m) for the case
where both $\ks$'s are used for the vertex reconstruction,
and 141$\mu$m (172$\mu$m) for the case where a single $\ks$ is used.
The latter resolution is comparable to the $f_{\rm tag}$ vertex position resolution.
The $B$ decay vertex in the tag side is determined from well-reconstructed tracks
that are not assigned to the $\bksks$ decay.
The typical vertex reconstruction efficiency for $f_{\rm tag}$ decays is determined to be 93$\%$.

\begin{figure}[htbp]
\begin{center}
\includegraphics[height=145pt,width=!]{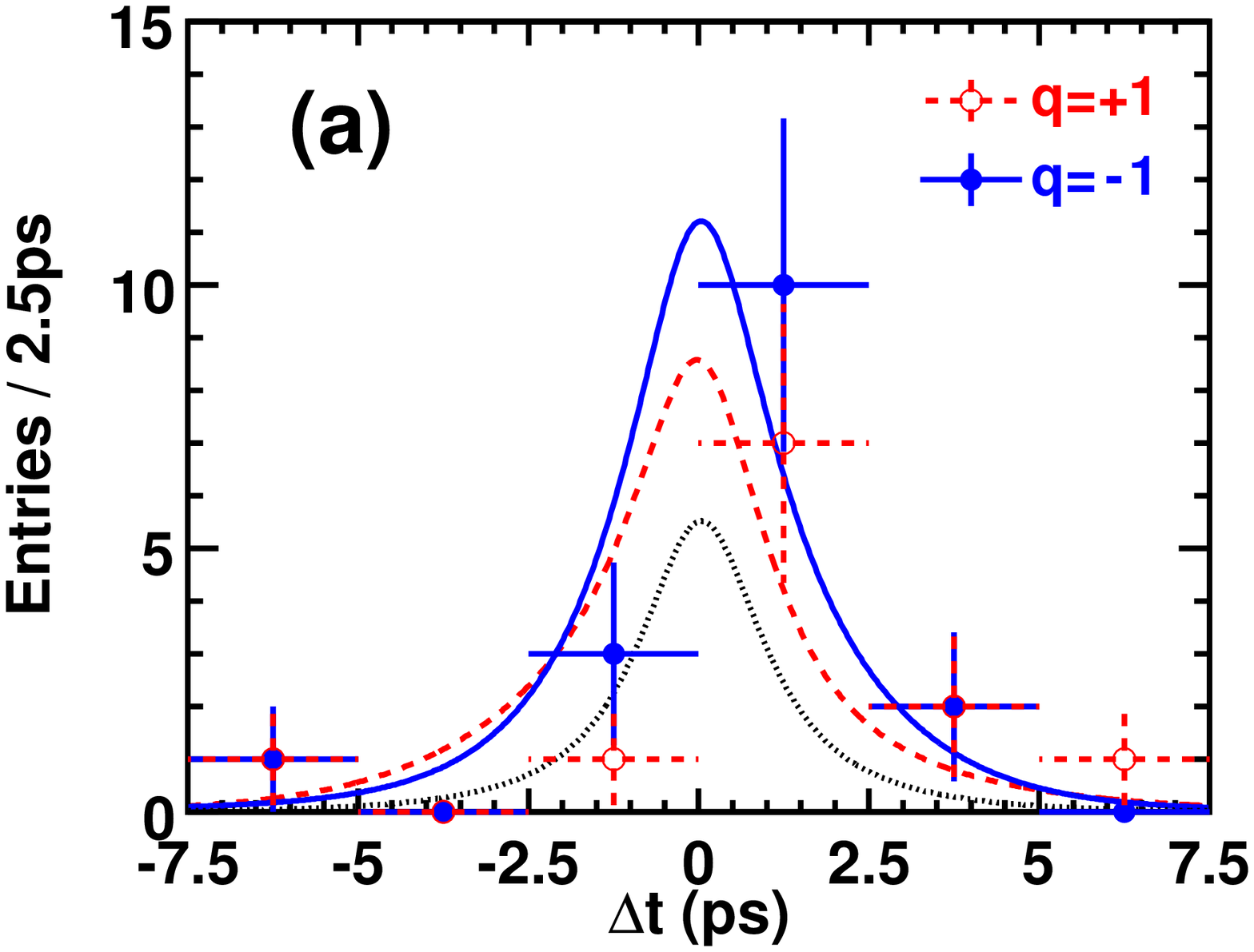} 
\includegraphics[height=145pt,width=!]{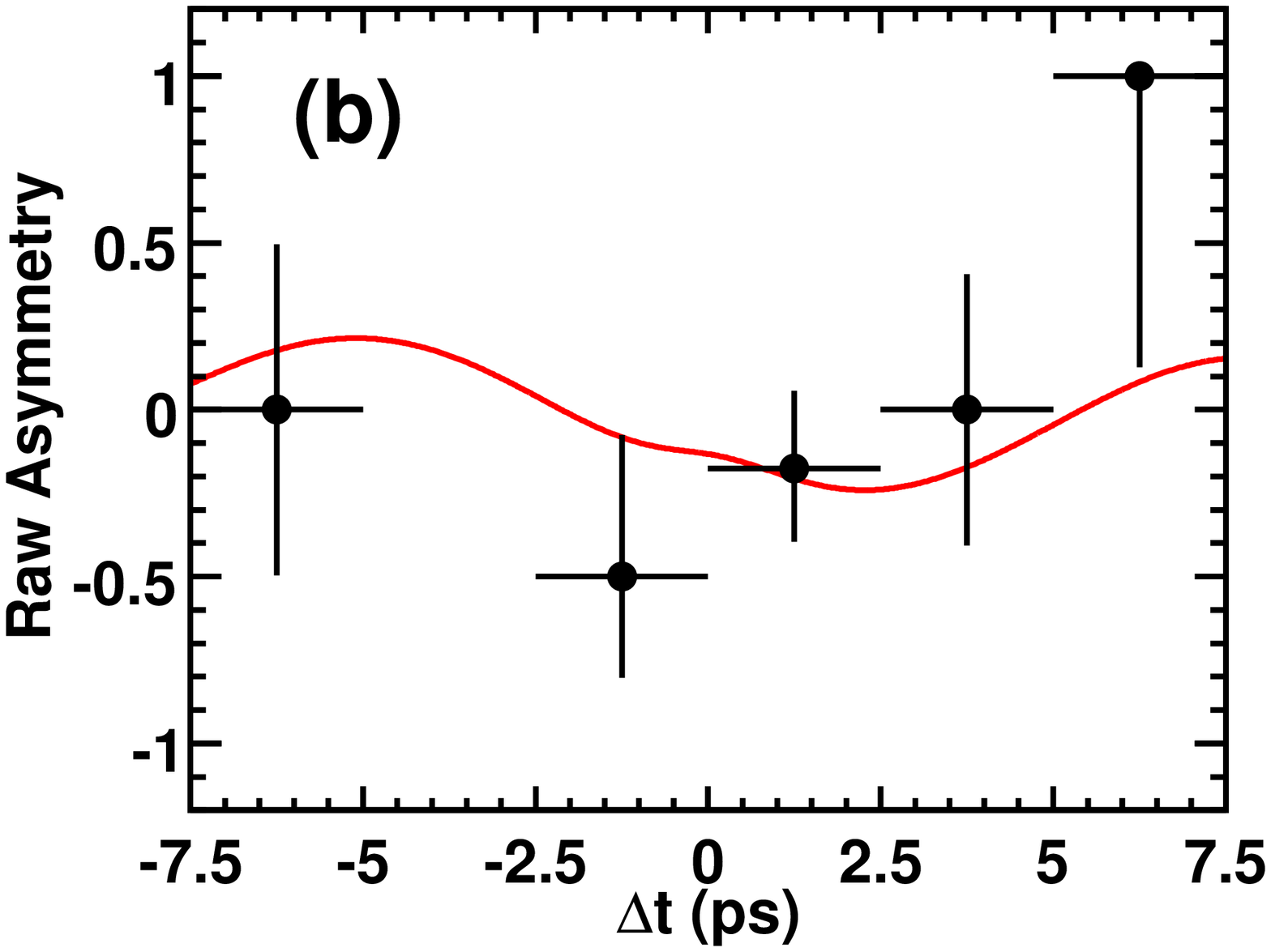} 
\caption{
(a) $\dt$ distribution and (b) raw asymmetry ${\cal A_{\rm CP}}$
for the $\bksks$ candidate events in the signal region with good tags ($r > 0.5$) and $\Lsb >$ 0.6.
In (a), the solid and dashed curves show the fit results with $q$ = $\pm$1, respectively.
The dotted curve shows the background component. In (b), the solid curve shows the fit projection.
}
\label{fig_dt}
\end{center}
\end{figure}

We determine $\sksks$ and $\aksks$ by performing an UML fit to the $\dt$ distribution.
For signal events, we use the $\dt$ distribution of Eq.~(\ref{eq_signal_pdf}),
modified to include the effect of incorrect flavor assignment.
The distribution ${\cal P}_{\ksks}$($\dt$) is then convolved with the resolution function $R_{\rm sig}(\dt)$,
which depends on the event-by-event vertex position errors~\cite{res-func};
the dependence is calibrated using a $B^0 \to J/ \psi \ks$ data control sample,
where the vertex positions are reconstructed using only a $\ks$ and the IP profile~\cite{ksvertex}.
We determine the following likelihood value for each event $i$:
\begin{eqnarray}
 P_i &=&
 (1-f_{\rm ol})\int [
   f_{\ksks}
  {\cal P}_{\ksks}(\dt^{\prime})
   R_{\rm sig}(\dt_i - \dt^{\prime}) \nonumber \\
 & &  +(1-f_{\ksks})
    {\cal P}_{\qq}(\dt^{\prime})
   R_{\qq}(\dt_i - \dt^{\prime})]
   d(\dt^{\prime}) \nonumber \\
 & & + f_{\rm ol}{\cal P}_{\rm ol}(\dt_i),
\label{eq_likelihood}
\end{eqnarray}
where the PDF ${\mathcal P_{\rm ol}}(\dt)$ is a broad Gaussian
that represents an outlier component with a small fraction $f_{\rm ol}$~\cite{res-func}.
The fraction $f_{\ksks}$ is the event-by-event signal fraction depending on $\de$, $\mbc$ and $\Lsb$.
We also take into account the $r$ dependence of the signal fraction $f_{\ksks}$;
we determine the dependence using the signal MC and the sideband events for the signal and background components,
respectively.
For background events, the $\dt$ distribution ${\cal P}_{\qq}(\dt)$ is convolved with a function $R_{\qq}(\dt)$,
where the distribution ${\cal P}_{\qq}(\dt)$ is modeled as a sum of an exponential function and a delta function,
and the function $R_{\qq}(\dt)$ is the sum of two Gaussians.
All parameters in ${\cal P}_{\qq}(\dt)$ and $R_{\qq}(\dt)$ are determined from sideband events.
We fix $\tau_{\bn}$ and $\Delta m_d$ to their world-average values~\cite{PDG}.
To improve the statistical sensitivity to $\aksks$,
we also use candidate events having no $\dt$ information, where both $\ks$'s decay outside the SVD
and we do not reconstruct $B$ vertices;
for these events, we use the PDF of Eq.~(\ref{eq_likelihood}) integrated over $\dt$.
The only free parameters in the fit are $\sksks$ and $\aksks$,
which are determined by maximizing the likelihood function $L$~=~$\prod {P}_i$,
where the product is over all events. The fit to 476 $\bksks$ candidate events,
in which 216 candidate events have no $\dt$ information, yields
\begin{eqnarray}
{\sksks} &=& \svalue ^{\sstaterrplus} _{\sstaterrminus} ({\rm stat}) \ssysterr {(\rm syst),} {\rm ~and}  \\
{\aksks} &=& \avalue \astaterr ({\rm stat}) \asysterr {(\rm syst),}
\end{eqnarray}
where the systematic errors are described below.
Figure~\ref{fig_dt} shows the $\dt$ distribution and raw asymmetry ${\cal A_{\rm CP}}$ in each $\dt$ interval,
where ${\cal A_{\rm CP}}$~=~($N_+ - N_{-}$)/($N_+ + N_{-}$), and $N_{\rm +(-)}$
is the number of candidate events with $q$~=~+1~($-$1).

The systematic error is primarily due to
uncertainties in the parameters of $R_{\rm sig}(\dt)$ ($\pm$0.06 on $\sksks$ and $<$0.01 on $\aksks$), and
uncertainties in the signal fraction $f_{\ksks}$ ($\pm$0.04 on $\sksks$ and $\pm$0.03 on $\aksks$).
We estimate a systematic error ($\pm$0.04 on $\sksks$ and $\pm$0.02 on $\aksks$)
for uncertainties in the parameters of ${\cal P}_{\qq}(\dt)$ and $R_{\qq}(\dt)$,
and the possible contribution of, and asymmetry in, the $B\overline{B}$ decay background.
The other contributions to the systematic errors come from uncertainties in
the wrong tag fraction ($\pm$0.02 on $\sksks$, $\pm$0.01 on $\aksks$),
fit biases ($\pm$0.02, $\pm$0.01),
physics parameters ($\tau_{\bn}$ and $\Delta m_d$) ($\pm$0.01, $\pm$0.01),
the vertex reconstruction ($\pm$0.01, $\pm$0.02),
and the tag-side interference effect~\cite{tsi} ($<$0.01, $\pm$0.03).
Adding all these contributions in quadrature,
we obtain systematic errors of $\ssysterr$ for $\sksks$ and $\asysterr$ for $\aksks$.

Various validity checks for the measurement are performed.
We measure a branching fraction for $B^0 \to K^0 \overline{K}{}^0$ of [$1.1\pm0.2$(stat)]$\times 10^{-6}$,
which is consistent with our previous measurement~\cite{ksksBelle}.
The $\bn$ lifetime for the $\bksks$ candidate events is measured to be 1.58$\pm$0.44 ps,
consistent with the world average value~\cite{PDG}.
We also fit to the sideband events of the $\bksks$ data sample and find no $CP$ asymmetry.
Using MC pseudo-experiments, we find that the statistical errors obtained in our measurement
are consistent with expectations.
We apply the same procedure to the $B^0 \to \jpsiks$ data sample without
using the $J / \psi$ daughter tracks for the vertex reconstruction.
We obtain $\scal _{\jpsiks} = 0.68 \pm 0.06$(stat),
which is in agreement with the world average for sin2$\phi _1$~\cite{PDG}.
We conclude that the vertex resolution for $\bksks$ decays is well-understood.
We reconstruct 1993$\pm$53 $\bkspc$~\cite{CC} events and,
without using the charged pion of the $B$ decay for vertex reconstruction,
apply the same fit procedure.
We obtain $\scal_{\ks \pi^{+}}=-0.13\pm 0.13$(stat)
and $\acal_{\ks \pi^{+}}=0.01\pm 0.06$(stat), which are consistent with no $CP$ asymmetry.

In summary, we measure time-dependent $CP$-violating parameters in $\bksks$ decays,
which are dominated by flavor-changing $b \to d\bar{s}s$ penguin transitions,
based on a data sample of $\totbb$$\times 10^6$ $\bbbar$ pairs recorded with the Belle detector.
We obtain $\sksks = \svalue ^{\sstaterrplus} _{\sstaterrminus} ({\rm stat})\ssysterr ({\rm syst})$
and $\aksks = \avalue \astaterr ({\rm stat})\asysterr ({\rm syst})$.
No $CP$ asymmetry is found for these decays.
These results are consistent with the SM prediction and also with the other measurement~\cite{ksksBabar}.
%

We thank the KEKB group for excellent operation of the
accelerator, the KEK cryogenics group for efficient solenoid
operations, and the KEK computer group and
the NII for valuable computing and Super-SINET network
support. We acknowledge support from MEXT and JSPS (Japan);
ARC and DEST (Australia); NSFC (China);
DST (India); MOEHRD, KOSEF and KRF (Korea);
KBN (Poland); MES and RFAAE (Russia); ARRS (Slovenia); SNSF (Switzerland);
NSC and MOE (Taiwan); and DOE (USA).

%


\begin{thebibliography}{99}
\bibitem{CC}
Throughout this paper,
the inclusion of the charge-conjugate decay mode is implied
unless otherwise stated.
\end{thebibliography}
\end{document}